\documentclass[12pt]{article}
\usepackage{amssymb}
  
\catcode`@=11
\def\secteqno{\@addtoreset{equation}{section}%
\def\theequation{\thesection.\arabic{equation}}}
\catcode`@=12
\secteqno
\newcommand{\be}{\begin{equation}}
\newcommand{\ee}{\end{equation}}
\newcommand{\bea}{\begin{eqnarray}}
\newcommand{\eea}{\end{eqnarray}}

\newcommand{\bref}[1]{(\ref{#1})}
\newcommand{\nn}{\nonumber}	   % fraction

\mathcode`\*="702A                  % * now always complex conjugate
\catcode163=13 \def\itm{\relax\ifmmode\to\else\itemize\fi}
\begin{document}
\thispagestyle{empty}
\hfill July 6, 2005

\hfill KEK-TH-1025

\vskip 20mm
\begin{center}
{\Large\bf Sugawara form for AdS superstring}
\vskip 6mm
\medskip
\vskip 10mm
{\large Machiko\ Hatsuda$^{\ast\dagger}$}

\parskip .15in
{\it $^\ast$Theory Division,\ High Energy Accelerator Research Organization (KEK),\\
\ Tsukuba,\ Ibaraki,\ 305-0801, Japan} \\{\it $^\dagger$Urawa University, Saitama \ 336-0974, Japan}\\
{\small e-mail:\ %\mhref
{mhatsuda@post.kek.jp}} \\
\parskip .35in

\medskip
\end{center}
\vskip 10mm
\begin{abstract}
We show that 
the stress-energy tensor for a superstring in the AdS$_5\times$S$^5$ background
is written in a supersymmetric generalized  ``Sugawara" form.
It is the ``supertrace" of the square of the right-invariant current
which is the Noether current satisfying the flatness condition.
The Wess-Zumino term is taken into account 
through the supersymmetric gauge connection in the right-invariant currents,
therefore the obtained stress-energy tensor is $\kappa$ invariant.
The integrability of the AdS superstring
provides
an infinite number of the conserved ``local" currents
which are  supertraces of the $n$-th power of the right-invariant current.
For even $n$ the ``local" current reduces to  terms proportional to the Virasoro
constraint and the $\kappa$ symmetry constraint,
while for 
odd $n$ it reduces to a term proportional to the $\kappa$ symmetry constraint .
\end{abstract} 

\setcounter{page}{1}
\parskip=7pt
\newpage
%%%%%%%%%%%%%%%%%%%%%%%%%%%%%%%%%%%%%%%%%%%%%%%%%%%%%%%%
\section{ Introduction}

The conjectured duality between
the type IIB superstring theory on the AdS$_5\times$S$^5$ space 
(AdS superstring)
and 
$D=4,~{\cal N}=4$ Yang-Mills theory
\cite{M,GKP,W}
has been driven not only 
studies of variety of background theories
but also studies of basic aspects such as integrability.
The approach of the pp-wave background superstring theory \cite{MTpp} 
was explored 
by Berenstein, Maldacena and Nastase \cite{BMN} and 
developed in, for example 
\cite{GKP2,FT}. 
For further development Mandal, Suryanarayan and Wadia 
pointed out the relevance with the integrability \cite{MSW},
and Bethe anzatz approach was explored 
 by Minahan and Zarembo \cite{MZ} and in for example \cite{B,DNW,BFST}.
The integrability 
is a powerful aspect expected in the large N QCD \cite{Lipatov}
and shown to exist in 
 the IIB superstring theory on the AdS$_5\times$S$^5$ space
by Bena, Polchinski and Roiban \cite{BPR}.
The integrability provides hidden symmetry generated by 
an infinite number of 
conserved ``non-local" charges \cite{LP,BIZZ}
as well as  
an infinite number of  conserved ``local" charges
\cite{pol2} which are related by a spectral parameter
at different points.
Related aspects on the integrability of the AdS superstring were
   discussed in \cite{new18}.

Recently the conformal symmetry of AdS superstrings
was conjectured due to  the $\kappa$ symmetry 
\cite{polyakov}.  
The classical conformal symmetry of the AdS superstring theory 
also  leads to an infinite number of conserved Virasoro operators.
The naive questions are
how the conformal generator  is related to the infinite number of conserved 
``local" currents, and how many independent conserved currents exist. 
For principal chiral models the stress-energy tensor 
is written by trace of the square of the conserved flat current;
for  reviews see refs.
\cite{EHMM,MSW}. 
For the AdS superstring theory 
the Wess-Zumino term and the $\kappa$ symmetry make a difference.
Recently issues related to
the integrability and the conformal symmetry 
of the AdS superstring theory have been discussed 
\cite{MP,Mh,AAT}.
In this paper we will obtain the expression of the conformal generator, 
which is the stress-energy tensor relating to 
 the lowest spin ``local" current,
  and we calculate the higher spin  ``local"  currents
to clarify independent components.

The AdS space contains the Ramond/Ramond flux which causes
difficulty of 
the standard Neveu-Schwarz-Ramond 
(NSR) formulation of the superstring theory.
The AdS superstring was described in the Green-Schwarz (GS) formalism
by Metsaev and Tseytlin based on the coset 
PSU(2,2$\mid$4)/[SO(4,1)$\times$SO(5)]
\cite{MT}.
Later  Roiban and Siegel reformulate it in terms of the unconstrained
GL(4$\mid$4) supermatrix coordinate based on an alternative 
coset GL(4$\mid$4)/[Sp(4)$\times$GL(1)]$^2$  \cite{RS}.
In this formalism the local Lorentz is gauged,
and it turns out that this treatment is essential for 
separation into  $+/-$ modes (right/left moving modes) easier.
Furthermore the fermionic constraint including the first class and second class
is necessary for
separation of the fermionic modes into $+/-$ modes.
As the first step toward the CFT formulation of the AdS superstring,
the affine Sugawara construction \cite{Halpern},
the Virasoro algebra and the algebra of currents carrying the 
space-time indices are also listed.

The organization of this paper is the following;
in the next section the notation is introduced.
In section 3 we analyze the superparticle in the AdS$_5\times$S$^5$ space,
and the relation between
the reparametrization constraint and the conserved right invariant (RI) current 
is given.
In section 4 we analyze the superstring in the AdS$_5\times$S$^5$ space,
and the infinite number of conserved currents are presented
both from the conformal point of view and
from the integrability point of view.
We show that the stress-energy tensor 
is written by
the ``supertrace" of the square of the RI current
 as the lowest spin ``local" current.
Then we  calculate higher spin ``local" currents   
to clarify independent components of the ``local" currents.

\par\vskip 6mm
%%%%%%%%%%%%%%%%%%%%%%%%%%%%%%%%%%%%%%%%%%%%%%%%%%%%%%%%%%%%%%%%%%%%%%%%%%%%%%%%%%%%%%%%%%%%%%%%%%%%%%%
\section{ GL(4${\mid}$4) covariant coset}

We review the Roiban-Siegel formulation of the  AdS$_5\times$S$^5$ coset
\cite{RS} and follow the notation in \cite{HKAdS}.  
The coset GL(4$\mid$4)/[GL(1)$\times$Sp(4)]$^2$ is used instead of 
PSU(2,2$\mid$4)/[SO(4,1)$\times$SO(5)] for the linear realization of the 
global symmetry after Wick rotations and introducing the auxiliary variables.
A coset element  $Z_M{}^A$
 is an unconstrained matrix defined on a world-volume 
 carrying indices $M=(m,\bar{m}),~A=(a,\bar{a})$ with 
 $m,\bar{m},a,\bar{a}=1,\cdots,4$.
The left invariant (LI) current, $L^L$, is invariant under the left action
$Z_M{}^A~\to~\Lambda_M{}^NZ_N{}^A
$ with
 a global parameter GL(4$\mid$4)$\ni \Lambda$
\bea
(J^L)_A{}^B=(Z^{-1}d Z)_A{}^B~~.
\eea
The LI current satisfies the flatness condition by definition
\bea
dJ^L=-J^LJ^L~~~.
\eea
The right  invariant (RI) current, $J^R$, is invariant under the right action 
$Z_M{}^A~\to~Z_M{}^B\lambda_B{}^A
$ with a local parameter 
[Sp(4)$\otimes$GL(1)]$^2$ $\ni\lambda$ 
\bea
(J^R)_M{}^N=({\cal D}ZZ^{-1})_M{}^N~~,~~({\cal D}Z)_M{}^A\equiv
dZ_M{}^A+Z_M{}^BA_B{}^A
\eea
with
\bea
A~\to~\lambda A\lambda^{-1}+(d\lambda) \lambda^{-1}~~,
\eea
and
\bea
dJ^R=J^RJ^R+Z(dA-AA)Z^{-1}~~~.\label{dAAA}
\eea
Originally $A$ is bosonic  [Sp(4)$\otimes$GL(1)]$^2$ 
$\ni$$A$, but we will show that 
the fermionic constraint i.e. $\kappa$ symmetry
gives fermionic components of $A$.

The conjugate momenta are introduced
\bea
\{
Z_M{}^A,\Pi_B{}^N
\}=(-)^A\delta_B^A\delta_M^N~~~
\eea
as the graded Poisson bracket and
 $\{q,p\}=-(-)^{qp}\{p,q\}$.
There are also two types of differential operators;
 the global symmetry generator (left action generator), $G_M{}^N$, 
and
the  supercovariant derivatives (right action generator), $D_A{}^B$, 
\bea
G_M{}^N=Z_M{}^A\Pi_A{}^N~~,~~D_A{}^B=\Pi_A{}^M Z_M{}^B~~~.\label{DGDG}
\eea
In our coset approach $8\times 8=64$ variables for $Z_M{}^A$ are introduced 
and auxiliary variables are eliminated by the following constraints
corresponding to the stability group [Sp(4)$\times $GL(1)]$^2$,
 \bea
({\bf D})_{(ab)}=(\bar{\bf D})_{(\bar{a}\bar{b})}={\rm tr}~{\bf D}=
{\rm tr}~\bar{\bf D}\equiv 0~~~\label{DSp4GL1}~~~,
\eea 
where the bosonic components are denoted by  boldfaced characters as
 ${\bf D}_{ab}\equiv D_{ab}$ and $\bar{\bf D}_{\bar{a}\bar{b}}\equiv
 D_{\bar{a}\bar{b}}$ of \bref{DGDG}.
The number of the coset constraints is $10+10+1+1=22$,
so the number of the coset parameters is $64-22=42$
where $10$ bosons and $32$ fermions.
The $[Sp(4)]^2$ invariant metric is anti-symmetric 
and a matrix is decomposed into
 trace part, anti-symmetric-traceless part and the symmetric part,
denoted by 
\bea
{\bf M}_{ab}=-\frac{1}{4}\Omega_{ab}{\bf M}^c{}_c+{\bf M}_{\langle ab\rangle}+{\bf M}_{(ab)}\equiv - \frac{1}{4}\Omega~{\rm tr}{\bf M}
+\langle {\bf M}\rangle+({\bf M})~~~,
\eea
with $M_{(ab)}=\frac{1}{2}(M_{ab}+M_{ba})$, 
and similar notation for the barred sector.

Both $G_M{}^N$ and $D_A{}^B$  in \bref{DGDG} satisfy GL(4$\mid$4) algebra.
If we focus on the AdS superalgebra part,
the global symmetry generators $G_M{}^N$ satisfies the global AdS superalgebra 
\bea
\left\{Q_{A\alpha},Q_{B,\beta}\right\}&=&-2\left[
\tau_3{}_{AB}P_{\alpha\beta} +\epsilon_{AB}
M_{\alpha\beta}\right]\label{QQPM}\\
Q_{1\alpha}&=&G_{m\bar{m}}+G_{\bar{m}m}%\cdots {\rm total~ supercharge}
\nn\\
Q_{2\alpha}&=&G_{m\bar{m}}-G_{\bar{m}m}%\cdots {\rm total~supercharge}
\nn\\
P_{\alpha\beta}
&=&{G}_{\langle mn\rangle}\Omega_{\bar{m}\bar{n}}
-G_{\langle\bar{m}\bar{n}\rangle}\Omega_{mn}\cdots {\rm total~ momentum}\nn\\
M_{\alpha\beta}&=&-G_{(mn)}\Omega_{\bar{m}\bar{n}}
+G_{(\bar{m}\bar{n})}\Omega_{mn}\cdots{\rm total~ Lorentz}\nn~~~.
\eea
The right hand side of \bref{QQPM} can not be diagonalized   
by the real SO(2) rotation of $Q_A$'s
because of the total Lorentz charge term with $\epsilon_{AB}$. 
On the other hand the local AdS supersymmetry algebra is given by
\bea
\left\{d_{A\alpha},d_{B,\beta}\right\}&=&2\left[
\tau_3{}
_{AB}
\tilde{p}_{\alpha\beta} +
\epsilon_{AB}m_{\alpha\beta}\right]\\
d_{1\alpha}&=&{D}_{a\bar{a}}+{\bar{D}}_{\bar{a}a}
\nn
\\
d_{2\alpha}&=&{D}_{a\bar{a}}-{\bar{D}}_{\bar{a}a}
\nn
\\
\tilde{p}_{\alpha\beta}&=&
{\bf D}
_{\langle ab\rangle}\Omega_{\bar{a}\bar{b}}
-{\bar{\bf D}}
_{\langle \bar{a}\bar{b}\rangle}\Omega_{ab}
\cdots {\rm local~LI~momentum}\nn
\\m_{\alpha\beta}&=&
-{\bf D}_{(ab)}\Omega_{\bar{a}\bar{b}}
+\bar{\bf D}
_{(\bar{a}\bar{b})}\Omega_{a{b}}\cdots{\rm local~Lorentz}\nn
~~~.
\eea 
In our coset approach the local Lorentz generator 
is a constraint \bref{DSp4GL1},
so  the 
local supercovariant derivative $d_{A\alpha}$'s
can be separated into;
\bea
\left\{d_{1\alpha},d_{2\beta}\right\}=
2 m_{\alpha\beta}\equiv 0~,~~
 \left\{d_{1\alpha},d_{1\beta}\right\}=
2 \tilde{p}_{\alpha\beta}~~,~~
\left\{d_{2\alpha},d_{2\beta}\right\}=
-2\tilde{p}_{\alpha\beta}
  \eea
Although the global superalgebra can not be separated into
irreducible algebras in the AdS background,
the local superalgebra can be separated into 
irreducible sets on the GL(4${\mid}$4) covariant coset approach. 
This property allows simpler description of the AdS superstring
as the flat case at least in the classical mechanics level. 

\par\vskip 6mm
%%%%%%%%%%%%%%%%%%%%%%%%%%%%%%%%%%%%%%%%%%%%%%%%%%%%%%%%%%%%%%%%%%%%%%%%%%%%%%%%%%%%%%%%%%%%%%%%%%%%%%%
\section{ AdS Superparticle}

We begin with the action for a superparticle in the AdS$_5\times$S$^5$ 
\bea
&S=\displaystyle\int  d\tau~\displaystyle\frac{1}{2e}
\left\{-{\bf J}_\tau^{\langle ab\rangle}{\bf J}_{\tau, \langle ab\rangle}
+\bar{\bf J}_\tau^{\langle \bar{a}\bar{b}\rangle}\bar{\bf J}_{\tau, \langle \bar{a}\bar{b}\rangle}
\right\}&~~~.\label{RS}
\eea
Here we omit the upper-subscript $L$ for the LI currents and their components are denoted as 
\bea
(J^L_{~\mu})_A{}^B
=\left(
\begin{array}{cc}
{\bf J}_{\mu,}{}_a{}^b&j_{\mu,}{}_a{}^{\bar{b}}\\\bar{j}_{\mu,}{}_{\bar{a}}{}^b&\bar{\bf J}_{\mu,}{}_{\bar{a}}{}^{\bar{b}}
\end{array}
\right)~~~.
\eea
From the definition of the canonical conjugates,
$
\Pi_A{}^M={\delta S}/{\delta \partial_\tau Z_M{}^A} (-)^A
$, 
we have
the following primary constraints \cite{HKAdS}
\bea
{\cal A}_{\rm P}=\frac{1}{2}{\rm tr}\left[
\langle{\bf D}\rangle^2-
\langle\bar{\bf D}\rangle^2\right]=0~~,~~
D_{a\bar{b}}=\bar{D}_{\bar{a}{b}}=0~~~\label{ApDD}
\eea
with 
\bea
D_A{}^B=\left(
\begin{array}{cc}{\bf D}_a{}^b&D_a{}^{\bar{b}}\\
\bar{D}_{\bar{a}}{}^b&\bar{\bf D}_{\bar{a}}{}^{\bar{b}}
\end{array}
\right)~~~.
\eea

The Hamiltonian is chosen as
\bea
{\cal H}=-{\cal A}_{\rm P}=- \frac{1}{2}{\rm tr}\left[
\langle{\bf D}\rangle^2-
\langle\bar{\bf D}\rangle^2\right]
\eea
and the $\tau$-derivative is determined by
the Poisson bracket with ${\cal H}$,
 $\partial_\tau{\cal O}=\{{\cal O},{\cal H}\}$. 
The fact that a half of the fermionic constraints  is second class 
requires the Dirac bracket in general.
Fortunately the Dirac bracket with the Hamiltonian is equal to 
its Poisson bracket because the fermionic 
constrains are ${\cal H}$ invariant.

The LI current is calculated as
\bea
J^L_\tau=Z^{-1}\partial_\tau Z=\left(
\begin{array}{cc}
\langle{\bf D}\rangle&0\\0&\langle \bar{\bf D}\rangle
\end{array}
\right)~~,~~\partial_\tau J^L=0~~~.
\eea
The RI current, generating the global GL(4$\mid$4) symmetry, 
is given as
\bea
J^R_\tau\equiv Z\Pi=Z\left(J^L_\tau+A_\tau\right)Z^{-1}~~,~~
A_\tau=
\left(
\begin{array}{cc}
({\bf D})-\frac{1}{4}\Omega {\rm tr}{\bf D}&D
\\\bar{D}&(\bar{\bf D})-\frac{1}{4}\Omega {\rm tr}\bar{\bf D}
\end{array}
\right)~~~.\label{JRSP}
\eea
Although the stability group does not contain fermionic components originally,
the fermionic components of the gauge connection $A$ in \bref{JRSP} is induced.
It is noted that ``$A$" is the gauge connection 
distinguishing from
 the reparametrization 
constraint ``${\cal A}$".
The RI current is conserved, since the Hamiltonian is written
by LI currents which are manifestly global symmetry invariant
\bea
\partial_\tau J^R=0~~~.
\eea

The $\kappa$ symmetry generators are half of the fermionic constraints
by projecting out with the null vector as
\bea
{\cal B}_{\rm P}{}_a{}^{\bar{b}}=\langle{\bf D}\rangle_a{}^b D_b{}^{\bar{b}}
+D_a{}^{\bar{a}}\langle\bar{\bf D}\rangle_{\bar{a}}{}^{\bar{b}}~~,~~
\bar{\cal B}_{\rm P}{}_{\bar{a}}{}^{{b}}=\langle\bar{\bf D}\rangle_{\bar{a}}{}^{\bar{b}}\bar{D}_{\bar{b}}{}^{{b}}
+\bar{D}_{\bar{a}}{}^{{a}}\langle{\bf D}\rangle_{a}{}^{b}~~~.
\eea
If we construct the closed algebra including these $\kappa$ generators 
with keeping
 the bilinear of the fermionic constraints, 
the $\tau$-reparametrization constraint, ${\cal A}_{\rm P}$, is modified to \cite{HKAdS}
\bea
\tilde{\cal A}_{\rm P}&=&\frac{1}{2}{\rm tr}\left[
\langle{\bf D}\rangle^2-
\langle\bar{\bf D}\rangle^2
+2D\bar{D}
\right]~~~~~.
\eea
This expression appears in the Poisson bracket of 
${\cal B}$ with $\bar{\cal B}$,
when we keep the bilinear of fermionic constrains.
The  RR flux is responsible for the last term ``$D\bar{D}$".
The term which is bilinear of the constraints 
does not change the Poisson bracket 
since its bracket with an arbitrary variable 
gives terms proportional to the constraints
which are zero on the constrained surface.
In another word ${\cal A}_{\rm P}$ has an ambiguity of bilinear of the constraints,
and the $\kappa$ invariance fixes it.
On the original coset constrained surface \bref{DSp4GL1} 
it is also rewritten as
\bea
\tilde{\cal A}_{\rm P}=\frac{1}{2}~{\rm Str}~[D_A{}^B]^2=\frac{1}{2}~{\rm Str}~[J^R_\tau]^2~~~.
\eea
This is zero-mode contribution of the classical Virasoro constraint
for a superstring
in the AdS$_5\times$S$^5$ background.

\par\vskip 6mm
%%%%%%%%%%%%%%%%%%%%%%%%%%%%%%%%%%%%%%%%%%%%%%%%%%%%%%%%%%%%%%%%%%%%%%%%%%%%%%%%%%%%%%%%%%%%%%%%%%%%%%%
\section{ AdS Superstring}
\subsection{Conserved currents}

We take the action for a superstring in the AdS$_5\times$S$^5$ given by
\bea
&S=\displaystyle\int  d^2\sigma~\frac{1}{2}\left\{
-\sqrt{-g}g^{\mu\nu}({\bf J}_\mu^{\langle ab\rangle}{\bf J}_{\nu, \langle ab\rangle}
-\bar{\bf J}_\mu^{\langle \bar{a}\bar{b}\rangle}\bar{\bf J}_{\nu, \langle \bar{a}\bar{b}\rangle}
)
+\frac{k}{2}\epsilon^{\mu\nu}(
E^{1/2}j_\mu^{a\bar{b}}j_{\nu, a\bar{b}}
-
E^{-1/2}\bar{j}_\mu^{\bar{a}{b}}\bar{j}_{\nu, \bar{a}{b}}
)\right\}&\nn\\\label{RS}
\eea
where ``$k$" represents the WZ term contribution with $k=1$ and
 $E={\rm sdet} Z_M{}^A$.
The consistent $\tau$ and $\sigma$ reparametrization generators \cite{HKAdS} are
\bea
{\cal A}_\perp&=&{\cal A}_{0\perp} +k~{\rm tr}\left[-E^{1/4}Fj_\sigma+E^{-1/4}\bar{F}\bar{j}_\sigma\right]\nn\\
{\cal A}_\parallel&=&{\cal A}_{0\parallel} +k~{\rm tr}\left[E^{-1/4}F\bar{j}_\sigma-E^{1/4}\bar{F}{j}_\sigma\right]\label{Aperp}
\eea
with the following primary constraints 
\bea
{\cal A}_{0\perp}&=&\frac{1}{2}{\rm tr}\left[
(\langle{\bf D}\rangle^2+\langle{\bf J}_\sigma\rangle^2)-
(\langle\bar{\bf D}\rangle^2+\langle\bar{\bf J}_\sigma\rangle^2)
\right]=0\nn\\
{\cal A}_{0\parallel}&=&{\rm tr}\left[
\langle{\bf D}\rangle\langle{\bf J}_\sigma\rangle-
\langle\bar{\bf D}\rangle\langle\bar{\bf J}_\sigma\rangle
\right]=0\\
F_{a\bar{b}}&=&E^{1/4}D_{a\bar{b}}
+\frac{k}{2}E^{-1/4}(\bar{j}_\sigma)_{\bar{b}a}=0\nn\\
\bar{F}_{\bar{a}{b}}&=&E^{-1/4}\bar{D}_{\bar{a}{b}}+\frac{k}{2}E^{1/4}({j}_\sigma)_{{b}\bar{a}}=0~~~.\label{FermionicF}
\eea
Their Poisson brackets are
\bea
\left\{{\cal A}_\perp(\sigma),{\cal A}_\perp(\sigma')\right\}&=&
2{\cal A}_\parallel(\sigma)\partial_\sigma\delta(\sigma-\sigma')+
\partial_\sigma{\cal A}_\parallel(\sigma)\delta(\sigma-\sigma')\nn\\
\left\{{\cal A}_\perp(\sigma),{\cal A}_\parallel(\sigma')\right\}&=&
2{\cal A}_\perp(\sigma)\partial_\sigma\delta(\sigma-\sigma')+
\partial_\sigma{\cal A}_\perp(\sigma)\delta(\sigma-\sigma')\\
\left\{{\cal A}_\parallel(\sigma),{\cal A}_\parallel(\sigma')\right\}&=&
2{\cal A}_\parallel(\sigma)\partial_\sigma\delta(\sigma-\sigma')+
\partial_\sigma{\cal A}_\parallel(\sigma)\delta(\sigma-\sigma')\nn~~~.
\eea

The Hamiltonian is chosen as
\bea
{\cal H}&=&-\int d\sigma {\cal A}_\perp\label{HamiltonianSUST}\\&=&
-\int d\sigma {\rm tr}\left[
\frac{1}{2}\left\{
\langle{\bf D}\rangle^2+\langle{\bf J}_\sigma\rangle^2-
\langle\bar{\bf D}\rangle^2-\langle\bar{\bf J}_\sigma\rangle^2\right\}
+\left(kE^{-1/2}\bar{D}\bar{j}_\sigma-k E^{1/2}Dj_\sigma
+j_\sigma\bar{j}_\sigma\right)
\right]
~~~.\nn
\eea
From now on $E=1$ gauge is taken using the local GL(1) invariance.
The global GL(1) symmetry is broken by the WZ term.

Using the Hamiltonian in  \bref{HamiltonianSUST}
the $\tau$-derivative of ${\cal A}_{\perp}$ and ${\cal A}_\parallel$ are given as
\bea
\partial_\tau {\cal A}_\perp=\partial_\sigma {\cal A}_\parallel~~,~~
\partial_\tau {\cal A}_\parallel=\partial_\sigma {\cal A}_\perp~~~.\label{dAdA}
\eea
Although the coset parameter $Z_M{}^A$ does not satisfy 
the world-sheet free wave equation, it is essential to introduce 
the world-sheet lightcone
coordinate 
\bea
\sigma^\pm=\tau \pm \sigma~~,~~
\partial_\pm=\frac{1}{2}(\partial_\tau \pm \partial_\sigma)~~~.
\eea
The differential equations \bref{dAdA} are rewritten as
\bea
\partial_- {\cal A}_+=0~~,~~
\partial_+ {\cal A}_-=0~~,~~
{\cal A}_\pm={\cal A}_\perp\pm {\cal A}_\parallel~~~,
\eea 
so the infinite number of the conserved 
currents are
\bea
\partial_- \left[f(\sigma^+){\cal A}_+\right]=0~~,~~
\partial_+ \left[f(\sigma^-){\cal A}_-\right]=0~~\label{conformal}
\eea
with an arbitrary function $f$.
Then there exist infinite number of conserved  charges
\bea
\partial_- \left[\displaystyle\int d\sigma~
f(\sigma^+){\cal A}_+\right]=0~~,~~
\partial_+ \left[\displaystyle\int d\sigma~
f(\sigma^-){\cal A}_-\right]=0~~~.
\eea

On the other hand the integrability of the superstring 
will provide the infinite number of ``local" charges as well as the 
``non-local" charges written down in \cite{HY}.
The LI currents
is given by 
\bea
\left\{\begin{array}{ccl}
J^L_\tau&=&\left(
\begin{array}{cc}
\langle{\bf D}\rangle&-k\bar{j}_\sigma\\-kj_\sigma&\langle \bar{\bf D}\rangle
\end{array}
\right)
=\left(
\begin{array}{cc}
\langle{\bf D}\rangle&
2D-2F\\2\bar{D}-2\bar{F}&\langle \bar{\bf D}\rangle
\end{array}
\right)
\approx
\left(
\begin{array}{cc}
\langle{\bf D}\rangle&
2D\\2\bar{D}&\langle \bar{\bf D}\rangle
\end{array}
\right)\nn\\
J^L_\sigma&=&
\left(
\begin{array}{cc}
{\bf J}_{\sigma}
&j_{\sigma}\\
\bar{j}_{\sigma}&
\bar{\bf J}_{\sigma}
\end{array}
\right)
\end{array}\right.~~~
\eea 
where the $\tau$ component is determined by \bref{HamiltonianSUST}. 
The LI currents satisfy the flatness condition but
does not satisfy the conservation law.
The RI currents are obtained in \cite{HY} as
\bea
\left\{\begin{array}{ccl}
J^R_\tau&=&ZDZ^{-1}=Z(J^L_\tau+A_\tau)Z^{-1}\label{RISUST}\\
J^R_\sigma&=&Z(J^L_\sigma+A_\sigma)Z^{-1}~~,~~
J^L_\sigma+A_\sigma=
\left(
\begin{array}{cc}
\langle{\bf J}_{\sigma}\rangle
&\bar{F}+\frac{1}{2}j_{\sigma}\\
F+\frac{1}{2}\bar{j}_{\sigma}&
\langle\bar{\bf J}_{\sigma}\rangle
\end{array}
\right)
\end{array}\right.
\eea
where the gauge connection $A_\mu$ is
\bea
\left\{\begin{array}{ccl}
A_\tau&=&\left(
\begin{array}{cc}
({\bf D})-\frac{1}{4}\Omega {\rm tr}{\bf D}&-D\\-\bar{D}&(\bar{\bf D})-\frac{1}{4}\Omega {\rm tr}\bar{\bf D}
\end{array}
\right)\nn\\
A_\sigma&=&
\left(
\begin{array}{cc}
-({\bf J}_\sigma)+\frac{1}{4}\Omega {\rm tr}{\bf J}_\sigma&
\bar{F}-\frac{1}{2}j_\sigma
\\F-\frac{1}{2}\bar{j}_\sigma
&-(\bar{\bf J}_\sigma)+\frac{1}{4}\Omega {\rm tr}\bar{\bf J}_\sigma
\end{array}
\right)
\end{array}\right.~~~.
\eea
The fermionic components of $A_\mu$ appear again.
In this paper
the fermionic constraints, $F$ and $\bar{F}$,
in 
the fermionic components of $A_\sigma$  are kept
while they were absent in our previous paper \cite{HY}
depending on the treatment of the constraint bilinear terms.
Then the integrability of the superstring 
leads to the current conservation and the flatness condition 
for the RI current;
\bea
\partial_\tau J^R_\tau=\partial_\sigma J^R_\sigma~~,~~
\partial_\tau J^R_\sigma-\partial_\sigma J^R_\tau=
2\left[J^R_\tau, J^R_\sigma\right]~~~\label{dJRdJR}~~~.
\eea
They are rewritten as
\bea
\partial_-J^R_+=\left[J^R_-,J^R_+\right]~~,~~
\partial_+J^R_-=\left[J^R_+,J^R_-\right]~~,~~
J^R_\pm=J^R_\tau\pm J^R_\sigma~~~.\label{JRpm}
\eea
Taking the supertrace, denoting ``Str", leads to 
the infinite number of conserved ``local"
currents because $J^R_\mu$ are supermatrices, 
\bea
\partial_-~{\rm Str}\left[(J^R_+)^n\right]=0~~,~~
\partial_+~{\rm Str}\left[(J^R_-)^n\right]=0~~,n=1,2,\cdots~~~.\label{JRn}
\label{integrablity}
\eea
It gives the infinite number of conserved ``local" charges
\bea
\partial_\tau \left[\displaystyle\int d\sigma~
 f(\sigma^+){\rm Str}(J^R_+)^n
\right]=0~~,~~
\partial_\tau \left[\displaystyle\int d\sigma~
 f(\sigma^-){\rm Str}(J^R_-)^n
\right]=0~~~.\eea

In this way classical 2-dimensional conformal symmetry and
integrability of AdS superstring lead to two infinite sets of 
 conserved currents, 
\bref{conformal} and \bref{integrablity}.
In next sections the relation between them is examined. 

%%%%%%%%%%%%%%%%%%%%%%%%%%%%%%%
\subsection{Stress-energy tensor ($n=2$)}

The ``$+/-$" (right/left moving) modes of the RI currents 
on the original coset constrained space 
\bref{DSp4GL1}
are written as
\bea
J^R_\pm=
%J^R_\tau \pm J^R_\sigma=
Z\left(
\begin{array}{cc}
\langle{\bf D}_\pm \rangle&D\pm (\bar{F}+\frac{1}{2}j_\sigma)
\\
\bar{D}\pm(F+\frac{1}{2}\bar{j}_\sigma)&\langle\bar{\bf D}_\pm\rangle
\end{array}
\right)Z^{-1}=
Z\left(
\begin{array}{cc}
\langle{\bf D}_\pm\rangle&d_\pm+\frac{1}{2}j_\pm
\\
\pm(d_\pm -\frac{1}{2}j_\pm)
&\langle\bar{\bf D}_\pm\rangle
\end{array}
\right)Z^{-1}\nn\\
\eea
  with
\bea
{\bf D}_\pm={\bf D}\pm{\bf J}_\sigma~~,~~
\bar{\bf D}_\pm=\bar{\bf D}\pm\bar{\bf J}_\sigma~~,~~d_\pm=F\pm\bar{F}~~,~~
j_\pm=j_\tau\pm j_\sigma=-\bar{j}_\sigma\pm j_\sigma
\label{pmpm}
\eea
carrying the LI currents indices, $AB$.
This is supertraceless, Str$J^R_\pm=0$, so $n=1$ case of
\bref{JRn} gives just trivial equation.

Let us look at the $n=2$ case of \bref{JRn},  
 ${\rm Str}\left[(J^R_\pm)^2\right]$.
Then the ``+" sector  is written as
\bea
\frac{1}{2}{\rm Str}\left[(J^R_+)^2\right]&=&
\frac{1}{2}{\rm Str}\left[
\left(
\begin{array}{cc}
\langle{\bf D}_+\rangle&d_++\frac{1}{2}j_+
\\
d_+-\frac{1}{2}j_+
&\langle\bar{\bf D}_+\rangle
\end{array}
\right)^2
\right]\nn\\
&=&\frac{1}{2}{\rm tr}\left[
\langle{\bf D}_+\rangle^2-\langle\bar{\bf D}_+\rangle^2
+2(d_++\frac{1}{2}j_+
)(d_+-\frac{1}{2}j_+
)
\right]\nn\\
&=&{\rm tr}\left[\frac{1}{2}\left(
\langle{\bf D}_+\rangle^2-\langle\bar{\bf D}_+\rangle^2\right)
+j_+d_+\right]~~~.\label{419}
\eea
The ``$-$" sector is
\bea
\frac{1}{2}{\rm Str}\left[(J^R_-)^2\right]&=&
{\rm tr}\left[\frac{1}{2}\left(
\langle{\bf D}_-\rangle^2-\langle\bar{\bf D}_-\rangle^2\right)
-j_-d_-
\right]~~~.\label{420}
\eea

On the other hand the conformal symmetry generator ${\cal A}_\pm$ is 
rewritten from the relation \bref{Aperp} and \bref{pmpm} as
\bea
{\cal A}_\pm&=&
{\rm tr} 
\left[\frac{1}{2}\left(
\langle{\bf D}_\pm \rangle^2-\langle\bar{\bf D}_\pm \rangle^2\right)
\pm j_\pm d_\pm
\right]~=~
\frac{1}{2}{\rm Str}\left[(J^R_\pm)^2\right]
~~~.\label{421}
\eea
If we take care of the square of the fermionic constraints,
the closure of the first class constraint set 
including the $\kappa$ symmetry 
generators, 
\bea
{\cal B}_\pm&=&
\langle{\bf D}_\pm\rangle d_\pm
+d_\pm\langle\bar{\bf D}_\pm\rangle\label{kappaSUST}
~~
\eea
determines the ambiguity of  bilinear of the constraints as
\bea
\tilde{\cal A}_\pm=
{\rm tr}\left[\frac{1}{2}\left(
\langle{\bf D}_\pm \rangle^2-\langle\bar{\bf D}_\pm \rangle^2\right)
\pm(\frac{1}{2}d_\mp +j_\pm)d_\pm
\right]=
{\cal A}_\pm+{\rm tr}F\bar{F}~~
\eea
obtained in \cite{HKAdS} as a generator of the ${\cal ABCD}$
constraint set
known to exist for a superstring in a flat space 
 \cite{WSmech,ABCD}. 
Then the stress-energy tensor is 
\bea
T_{\pm\pm}\equiv\tilde{\cal A}_\pm
\approx
{\cal A}_\pm
={\rm Str}J^R_\pm J^R_\pm 
~~~.\label{423}
\eea
This is $\kappa$ symmetric stress-energy tensor 
in a supersymmetric generalization of
 Sugawara form.

%%%%%%%%%%%%%%%%%%%%%%%%%%%%%%%%%%%%%%%%%%%%%%%%%%%%%%%%%%%%%%%%%%%%%%%%%%%%%%%%%%%%%%%%%%%%%%%%%%%%%%

\subsection{Supercovariant derivative algebra}

Existence of the conformal invariance should present the 
irreducible coset components of supercovariant derivatives
\cite{HKAdS};
\bea
\langle{\bf D}_\pm\rangle&=&\langle{\bf D}\rangle\pm\langle{\bf J}_\sigma\rangle
~~,~~
\langle\bar{\bf D}_\pm\rangle~=~\langle\bar{\bf D}\rangle\pm\langle\bar{\bf J}_\sigma\rangle\nn\\
d_\pm&=&F\pm\bar{F}=(D\pm\frac{1}{2}j_\sigma)\pm(\bar{D}\pm\frac{1}{2}\bar{j}_\sigma)
\nn~~~.
\eea

On the  constraint surface \bref{DSp4GL1} and \bref{FermionicF}
the $+/-$ sector supercovariant derivatives are separated as
\bea
\left\{\langle{\bf D}_+\rangle_{ab}(\sigma),
\langle{\bf D}_-\rangle_{cd}(\sigma')\right\}
&=&2\Omega_{\langle c|\langle b}({\bf D})_{a\rangle|d\rangle}
\delta(\sigma-\sigma')
\equiv 0\nn\\
\left\{\langle{\bf D}_+\rangle_{ab}(\sigma),
d_{-,c\bar{d}}(\sigma')\right\}
&=&\Omega_{c\langle b}d_{+,a\rangle \bar{d}}
\delta(\sigma-\sigma')=
\Omega_{c\langle b}(F+\bar{F})_{a\rangle \bar{d}}
\delta(\sigma-\sigma')
\approx 0\nn\\
\left\{d_{+,a\bar{b}}(\sigma),
d_{-,c\bar{d}}(\sigma')\right\}
&=&2\left[
\Omega_{ac}(\bar{\bf D})_{\bar{b}\bar{d}}
+\Omega_{\bar{b}\bar{d}}({\bf D})_{ac}
\right]
\delta(\sigma-\sigma')
\equiv 0\nn
\eea
with 
analogous relation for  the barred sector, $\langle\bar{\bf D}_\pm\rangle$.

The ``+" sector supercovariant derivative algebra is
\bea
\left\{\langle{\bf D}_+\rangle_{ab}(\sigma),
\langle{\bf D}_+\rangle_{cd}(\sigma')\right\}
&=&2\Omega_{\langle c|\langle b}\Omega_{a\rangle|d\rangle}
\delta'(\sigma-\sigma')+4\Omega_{\langle c|\langle b}
({\bf J}_\sigma)_{a\rangle|d\rangle}
\delta(\sigma-\sigma')\nn\\
&\equiv& 2\Omega_{\langle c|\langle b} \nabla_{a\rangle|d\rangle}
\delta(\sigma-\sigma')
\nn\\
\left\{d_{+,a\bar{b}}(\sigma),
d_{+,c\bar{d}}(\sigma')\right\}
&=&2\left[
\Omega_{\bar{b}\bar{d}}\langle{\bf D}_+\rangle_{ac}
-\Omega_{ac}\langle\bar{\bf D}_+\rangle_{\bar{b}\bar{d}}
\right]
\delta(\sigma-\sigma')
\nn\\
\left\{\langle{\bf D}_+\rangle_{ab}(\sigma),
d_{+,c\bar{d}}(\sigma')\right\}
&=&
\Omega_{c\langle b}(d_-+2j_+)_{a\rangle \bar{d}}
\delta(\sigma-\sigma')\approx
2\Omega_{c\langle b}\omega_{+,a\rangle \bar{d}}
\delta(\sigma-\sigma')
\nn\\
\left\{d_{+,a\bar{b}}(\sigma),
\omega_{+,c\bar{d}}(\sigma')\right\}
&=&-2\Omega_{\bar{b}\bar{d}}\Omega_{ac}\delta'(\sigma-\sigma')
2\left[
-\Omega_{\bar{b}\bar{d}}({\bf J}_\sigma)_{ac}
-\Omega_{ac}(\bar{\bf J}_\sigma)_{\bar{b}\bar{d}}\right]\delta(\sigma-\sigma')\nn\\
&\equiv& -2\nabla_{\bar{b}\bar{d};ac}\delta(\sigma-\sigma')\nn\\
\label{sucovder}\\
\left\{\langle{\bf D}_+\rangle_{ab}(\sigma),
\omega_{+,c\bar{d}}(\sigma')\right\}
&=&
\Omega_{c\langle b}\omega_{-,a\rangle \bar{d}}
\delta(\sigma-\sigma')
\nn\\
\left\{\omega_{+,a\bar{b}}(\sigma),
\omega_{+,c\bar{d}}(\sigma')\right\}
&=&0~~~\nn
\eea
where
\bea
\omega_\pm&=&j_\pm=-\bar{j}_\sigma\pm j_\sigma~~~.
\eea
This is comparable with the flat case where
the non-local term, $\partial_\sigma\delta(\sigma-\sigma')$, is replaced by the 
local Lorentz (~[Sp(4)]$^2$~) covariant non-local term, 
$\nabla_\sigma\delta(\sigma-\sigma')$. 
For the fifth Poisson bracket,
 $\left\{\langle{\bf D}_+\rangle,
\omega\right\}
$, 
it is zero for the flat case but it is not for the AdS case.
For a superstring in a flat space 
the consistency of the $\kappa$ symmetry constraint
requires 
the first class constraint set,
namely ``${\cal ABCD}$" constraint,
which are bilinear of the supercovariant derivatives
 \cite{WSmech,ABCD}. 
 For the AdS case the situation is completely the same,
despite of this anomalous term \cite{HKAdS}.

\par\vskip 6mm

%%%%%%%%%%%%%%%%%%%%%%%%%%%%%%%%%%%%%%%%%%%%%%%%%%%%%%%%%%%%%%%%%%%%%%%%%%%%%%%%%%%%%%%%%%%%%%%%%%%%%%

\subsection{``Local" currents ($n\geq 3$)}

Next let us look at $n\geq 3 $ cases
of
the infinite number of conserved ``local" current \bref{JRn}.
For simplicity we focus on the ``+" sector and replace 
$``+"$ by $``~\hat{~}~"$, as
$J_+ \to \hat{J}$.
The first  three powers of the RI current, 
$(J^R)^n$ with $n=1,2,3$, are listed as below:
\bea
\left[Z^{-1}\hat{J}^R Z\right]_{AB}&=&\left(
\begin{array}{cc}
\langle\hat{\bf D}\rangle_{\langle ab\rangle}&
(\hat{d}+\frac{1}{2}\hat{j})_{a\bar{b}}
\\
\pm(\hat{d} -\frac{1}{2}\hat{j})_{b\bar{a}}
&\langle\hat{\bar{\bf D}}\rangle_{\bar{a}\bar{b}}
\end{array}
\right)
\eea
%%%%%%%%%%%
\bea
\left[Z^{-1}(\hat{J}^R)^2 Z\right]_{AB}&=&-\frac{1}{4}
\left(
\begin{array}{cc}
\Omega_{ab}~{\rm tr}(
\langle\hat{\bf D}\rangle^2
+\hat{j}\hat{d})
&\\&
\Omega_{\bar{a}\bar{b}}~{\rm tr}(
\langle\hat{\bar{\bf D}}\rangle^2
-\hat{j}\hat{d})
\end{array}
\right)\\
&&+
\left(
\begin{array}{cc}
(\hat{d}^2-\frac{1}{4}\hat{j}^2)_{(ab)}
+\langle \hat{j}\hat{d}\rangle_{\langle ab \rangle}
&\hat{\cal B}_{a\bar{b}}
+\frac{1}{2}(\langle \hat{\bf D}\rangle \hat{j}
+\hat{j}\langle \hat{\bar{\bf D}}\rangle
)_{a\bar{b}}
\\
\hat{\cal B}_{b\bar{a}}
-\frac{1}{2}(\langle \hat{\bf D}\rangle \hat{j}
+\hat{j}\langle \hat{\bar{\bf D}}\rangle
)_{b\bar{a}}
&(\hat{d}^2-\frac{1}{4}\hat{j}^2)_{(\bar{a}\bar{b})}
-\langle\hat{j}\hat{d}\rangle_{\langle \bar{a}\bar{b} \rangle}
\end{array}
\right)\nn
\eea
%%%%%%%%%%%
\bea
&&\left[Z^{-1}(\hat{J}^R)^3 Z\right]_{AB}~=~
\frac{1}{4}
\left(
\begin{array}{cc}
\Omega_{ab}~{\rm tr}\left[\hat{\cal B}\hat{j}-
(\langle\hat{\bf D}\rangle \hat{d} )\hat{j}
\right]
&\\&
-\Omega_{\bar{a}\bar{b}}~{\rm tr}\left[\hat{\cal B}\hat{j}
-(\hat{d}\langle\hat{\bar{\bf D}}\rangle)\hat{j}
\right]
\end{array}
\right)\\
&&-
\left(
\begin{array}{cc}
\left[\frac{1}{4}{\rm tr}(\langle\hat{\bf D}\rangle^2+\hat{j}\hat{d})~
\langle\hat{\bf D}\rangle
-\langle\hat{\bf D}\rangle (\hat{j} \hat{d})
+\hat{\cal B}\hat{j}
\right]_{\langle ab\rangle}
&
\frac{1}{4}{\rm tr}(\langle\hat{\bf D}\rangle^2
-\langle\hat{\bar{\bf D}}\rangle^2)~(\hat{d}+\frac{1}{2}\hat{j})_{a\bar{b}}
\\
\frac{1}{4}{\rm tr}(\langle\hat{\bf D}\rangle^2
-\langle\hat{\bar{\bf D}}\rangle^2)~(\hat{d}-\frac{1}{2}\hat{j})_{b\bar{a}}
&
\left[\frac{1}{4}{\rm tr}(\langle\hat{\bar{\bf D}}\rangle^2
+\hat{j}\hat{d})~
\langle\hat{\bar{\bf D}}\rangle
-(\hat{j} \hat{d})
\langle\hat{\bar{\bf D}}\rangle -\hat{\cal B}\hat{j}
\right]_{\langle \bar{a}\bar{b}\rangle}
\end{array}
\right)\nn\\
&&+
\left(
\begin{array}{cc}
\left[
2(\hat{d}^2-\frac{1}{4}\hat{j}^2)\langle\hat{\bf D}\rangle
+\hat{d}\langle\hat{\bar{\bf D}}\rangle\hat{d}
-\frac{1}{4}\hat{j}\langle\hat{\bar{\bf D}}\rangle \hat{j}
\right]_{(ab)}&
-\frac{1}{4}{\rm tr}(\hat{j}\hat{d})
(\hat{d}+\frac{1}{2}\hat{j})_{a\bar{b}}
+
\left[\langle\hat{{\bf D}}\rangle
(\hat{d}+\frac{1}{2}\hat{j})\langle\hat{\bar{\bf D}}\rangle
\right]_{a\bar{b}}
\\
\frac{1}{4}{\rm tr}(\hat{j}\hat{d})
(\hat{d}-\frac{1}{2}\hat{j})_{b\bar{a}}
+
\left[\langle
\hat{{\bf D}}\rangle
(\hat{d}-\frac{1}{2}\hat{j})\langle\hat{\bar{\bf D}}\rangle
\right]_{b\bar{a}}
&
\left[
2(\hat{d}^2-\frac{1}{4}\hat{j}^2)\langle\hat{\bar{\bf D}}\rangle
+\hat{d}\langle\hat{{\bf D}}\rangle\hat{d}
-\frac{1}{4}\hat{j}\langle\hat{{\bf D}}\rangle \hat{j}
\right]_{(\bar{a}\bar{b})}
\end{array}
\right)\nn\\
&&+
\left(
\begin{array}{cc}
&
\left[\left\{
(\hat{d}^2-\frac{1}{4}\hat{j}^2)
+\langle\hat{j}\hat{d}
\rangle
\right\}(\hat{d}+\frac{1}{2}\hat{j})
\right]_{a\bar{b}}
\\
\left[\left\{
(\hat{d}^2-\frac{1}{4}\hat{j}^2)
-\langle\hat{j}\hat{d}
\rangle
\right\}(\hat{d}-\frac{1}{2}\hat{j})
\right]_{\bar{a}b}
&
\end{array}
\right)\nn
\eea
In this computation
 5-dimensional $\gamma$-matrix relations are used,
for example 
${\bf V}^{\langle ab\rangle}{\bf U}_{\langle bc\rangle}
+{\bf U}^{\langle ab\rangle}{\bf V}_{\langle bc\rangle}
=\frac{1}{2}\delta^a_c~{\rm tr}{\bf V}{\bf U}$ for bosonic
vectors ${\bf V},~{\bf U}$.

The conserved ``local" current with $n=3$ becomes
\bea
{\rm Str}
(\hat{J}^R)^3 &=&
{\rm tr}\left[2\hat{\cal B}\hat{j}
-(\langle\hat{\bf D}\rangle \hat{d} )\hat{j}
-(\hat{d}\langle\hat{\bar{\bf D}}\rangle)\hat{j}
\right]
=~{\rm tr}
~(\hat{\cal B}\hat{j}) \label{430}
\eea
where $\hat{\cal B}$ is the $\kappa$ generating constraint
 \bref{kappaSUST}.
 The conserved ``local" current with $n=4$ becomes
\bea
{\rm Str}
(\hat{J}^R)^4 &=&
-\frac{1}{2}
{\rm tr}\left(
\langle\hat{\bf D}\rangle^2+
\langle\hat{\bar{\bf D}}\rangle^2
\right)
\hat{\cal A}+\left(~\cdots~\right){\rm tr}
~(\hat{\cal B}\hat{j})~~.
\eea
The conserved ``local" current with $n=5,6$ are given as; 
 Str$(\hat{J}^R)^5=$(
$\hat{\cal B}$ dependent terms), 
 Str$(\hat{J}^R)^6=$(
$\hat{\cal A}$ and $\hat{\cal B}$ dependent terms). 
In general for even $n=2m$ its bosonic part is given as
\bea
{\rm Str}
(\hat{J}^R)^{2m} \mid_{\rm bosonic}&=&
\left({\rm tr}\langle\hat{\bf D}\rangle^2\right)^m-
\left({\rm tr}\langle\hat{\bar{\bf D}}\rangle^2\right)^m\nn\\
&=&{\rm tr}\left(\langle\hat{\bf D}\rangle^2-\langle\hat{\bar{\bf D}}\rangle^2
\right)~\left\{\left(
{\rm tr}\langle\hat{\bf D}\rangle^2\right)^{m-1}
+\cdots +
\left({\rm tr}\langle\hat{\bar{\bf D}}\rangle^2\right)^{m-1}
\right\}\nn\\
&\Rightarrow& (\cdots)\hat{\cal A}+\left(~\cdots~\right){\rm tr}
~(\hat{\cal B}\hat{j})
\eea
where the last equality is guaranteed by the $\kappa$ invariance.
It is also pointed out that
the conserved supertraces of multilinears in the currents factorize in traces of lower number of currents and that for an even number of currents one of the factors
is the stress tensor in \cite{BPR}. 
For odd $n=2m+1$ its bosonic part is given as
\bea
{\rm Str}
(\hat{J}^R)^{2m+1} \mid_{\rm bosonic}~=~0
~\Rightarrow~\left(~\cdots~\right){\rm tr}
~(\hat{\cal B}\hat{j})
\eea
where the possible fermionic variable dependence is
a term proportional to $\hat{\cal B}$
 guaranteed by the $\kappa$ invariance.

In this way, after taking supertrace the even 
$n$-th power of $J^R$ 
reduces terms proportional to ${\cal A}$ and ${\cal B}$,
and the odd   $n$-th power of $J^R$ 
reduces a term proportional to ${\cal B}$ only.
In this paper
${\cal CD}$ constraints in the ${\cal ABCD}$ first class constraint set 
are not introduced for simpler argument,
and set to zero because they are bilinears of constraints.

\par\vskip 6mm
%%%%%%%%%%%%%%%%%%%%%%%%%%%%%%%%%%%%%%%%%%%%%%%%%%%%%%%%%%%%%%%%%%%%%%%%%%%%%%%%%%%%%%%%%%%%%%%%%%%%%%%
\section{ Conclusion and discussions}

We obtained
the expression of 
the conserved ``local" currents 
derived from the integrability of a superstring
in the AdS$_5\times$S$^5$ background.
The infinite number of the conserved ``local" currents
 are written by the supertrace of the $n$-th power of the RI currents.
 The lowest nontrivial case, $n=2$, is nothing but the stress-energy tensor
  which is also Virasoro constraint, 
  Str$(J^R)^2_\pm$ in \bref{419} and \bref{420}.
  For even $n$ the ``local" current reduces to terms proportional to the 
  Virasoro constraint and the $\kappa$ symmetry constraint.
 For odd $n$ it reduces to a term   proportional to the
 $\kappa$ symmetry constraint.
In another word  the integrability
reduces to 
the ${\cal AB}({\cal CD})$ first class constraint set
where ${\cal A}$ is the Virasoro generator
and ${\cal B}$ is the $\kappa$ symmetry generator.
The ${\cal ABCD}$ first class constraint set 
is the local symmetry generator of superstrings both on the flat space
and on the AdS space. 
It is natural that the physical degrees of freedom of 
a superstring is common locally,
  independently of  flat or AdS backgrounds.

 It seems that the combination of the  ${\cal B}_\pm j_\pm$ 
 in \bref{430} 
plays the role of the world-sheet supersymmetry operator
in a sense of the grading of the conformal generator.
However it is not
straightforward
to construct the worldsheet supersymmetry operator.
As in the flat case where the lightcone gauge makes 
the relation between 
the GS fermion and the NSR fermion more transparent,
the $\kappa$ gauge fixing will be a clue to 
make a connection to the world-sheet supersymmetry. 
We leave this problem in addition to the quantization 
problem for future investigations.

\par\vskip 6mm
%%%%%%%%%%%%%%%%%%%%%%%%%%%%%%%%%%%%%%%%%%%%%%%%%%%%%%%%%%%%%%%%%%%%%%%%%%%%%%%%%%%%%%%%%%%%%%%%%%%%%%%

\noindent{\bf Acknowledgments}

The author thanks to K. Kamimura, S. Mizoguchi and K. Yoshida for fruitful discussions.

\par\vskip 6mm
%%%%%%%%%%%%%%%%%%%%%%%%%%%%%%%%%%%%%%%%%%%%%%%%%%%%%%%%%%%%%%%%%%%%%%%%%%%%%%%%%%%%%%%%%%%%%%%%%%%%%%%
\appendix

%\section*{Appendix}

%\section{Notation} 

%%%%%%%%%%%%%%%%%%%%%%%%%%%%%%%%%%%%%%%%%%%%%%%%%

\end{document}